# Silver paint as a soldering agent for DyBaCuO single-domains welding

J-P. Mathieu[1], J-F. Fagnard[2], Ph. Laurent[2], B. Mattivi[2], C. Henrist[3], Ph. Vanderbemden[2], M. Ausloos[4], R. Cloots[1,2,3]

[1] S.U.P.R.A.T.E.C.S., Chemistry Institute B6,.University of Liege, Sart-Tilman, 4000 Liege, Belgium
[2] S.U.P.R.A.T.E.C.S., Department of Electrical Engineering and Computer Science B28, University of Liege, Sart-Tilman, 4000 Liege, Belgium
[3] CAT , University of Liege, Sart-Tilman B6, B-4000 Liege, Belgium (www.catmu.ulg.ac.be )
[4] S.U.P.R.A.T.E.C.S., Physics Institute B5,.University of Liege, Sart-Tilman, 4000 Liege, Belgium

ABSTRACT:

Silver paint has been tested as a soldering agent for DyBaCuO single-domain welding. Junctions have been manufactured on Dy-Ba-Cu-O single-domains cut either along planes parallel to the *c*-axis or along the *ab*-planes. Microstructural and superconducting characterisations of the samples have been performed. For both types of junctions, the microstructure in the joined area is very clean: no secondary phase or Ag particles segregation has been observed. Electrical and magnetic measurements for all configurations of interest are reported ($\rho$(T) curves, and Hall probe mapping). The narrow resistive superconducting transition reported for all configurations shows that the artificial junction does not affect significantly the measured superconducting properties of the material.

INTRODUCTION

Nowadays, the production of a large number of good quality (RE)BaCuO (RE = rare earth) single-domain bulk samples by the top-seeded melt textured growth (TSMTG) process is well established [1]. Nevertheless, applications based on superconducting technology require complex-shaped superconducting single-domains. For example, a superconducting ring is required for making flywheels [2-4]. More complex shapes are required for applications involving superconducting rotating machines [5] or magnetic field shields [6].

In order to produce complex-shaped samples, the growth of large single grains which would be drilled and machined is not the best solution [7]. The main reason is that the sample inhomogeneity increases with increasing size. These inhomogeneities mainly affect the transport current; consequently large samples usually display a smaller critical current density, $J_c$, than smaller ones. Moreover, machining ceramics, and especially superconducting (RE)BaCuO single-domains, is very difficult because of the poor mechanical properties of such materials; they are very brittle [8].

An alternative way to produce complex shape superconducting materials is to use the Infiltration and Growth (IG) process instead of TSMTG process. The IG process is based on the infiltration of a barium and copper rich liquid phase into a RE-211 preform, which could have a complex shape, even a foam structure, at a temperature near the peritectic temperature of the RE-123 phase [7,9-17]. This technique allows obtaining high density materials with small size RE-211-particles well dispersed in the bulk [9]. Moreover a near-net shape material can be obtained. Indeed, the process leads to a limited shrinkage which reduces cracks and distortions in the superconducting matrix

In order to manufacture large single-domains, it is also possible and efficient to weld small high-quality single-domains together. In such a case, the junction has to display the highest possible critical current density. Various techniques for joining (RE)BaCuO single-domains have been developed: *(i)* a natural joining using a multi-seeding technique [18-23] ; *(ii)* diffusion bonding of polished surfaces under uniaxial pressure in absence of soldering agents; and *(iii)* artificial joining by using low peritectic temperature rare earth cuprate ceramics, like $YbBa_2Cu_3O_7$, $ErBa_2Cu_3O_7$ or $TmBa_2Cu_3O_7$, as soldering agents [24-29]. Silver has been also used to reduce locally the peritectic temperature of the RE-123 phase [20,30,31]. Unfortunately, using low melting point RE-123 compounds leads to microstructural inhomogeneities which affect locally the critical current density. YBCO/Ag used as soldering agent gives better results. However the processing time is long because YBCO/Ag single-domains have to be prepared prior to manufacture the junction.

Recently Iliescu *et al.* [32] have proposed an alternative method for single-domain welding along (100) or (010) planes. The authors suggest the use of a silver foil as a welding agent. This set-up gives good results and allows scaling up the production of welded samples.

In this paper, we have studied the efficiency of silver paint as a soldering agent. The main advantage of silver paint compared with the silver foil is a greatly simplified processing. Pieces to be welded quickly stick together after the evaporation of the solvent. The manipulation of the samples is thus made easier. More precisely we have constructed junctions of DyBaCuO single-domains either along the (100) or (010) planes (planes parallel to the *c*-axis) or (001) planes (planes perpendicular to the *c*-axis); see figure 1. These junctions have been characterised by microscopy and electrical measurements; all configurations of interest are reported.

EXPERIMENTAL

DyBaCuO single-domains were at first produced by a Top-Seeded Melt-Textured Growth process using a $Sm_{1.8}Ba_{2.4}Cu_{2.4}O_x$ single-grain seed. The details about the single-domain synthesis are described elsewhere [8].

Single-domain samples were cut along a plane parallel to the *c*-axis or along an *ab*-plane (plane perpendicular to the *c*-axis) with a diamond disk fretsaw. The so-obtained faces were finely polished in order to obtain flat surfaces for welding.

A silver paint, consisting of silver particle suspension in iso-butyl methyl ketone solvent, was used to cover the faces to join. These faces were quickly manually pressed against each other before the solvent evaporated. Since the silver paint makes the different pieces sticking together, complex-shape samples with multi-junctions can be easily constructed. The thickness of the silver layer was optically observed to be about 50 to 60 µm depending on the painting conditions.

The sample was then subjected to a heat treatment. The sample was first heated rapidly to a temperature of 990°C and maintained at this temperature for 1 h. This step was followed by a slow cooling ramp (0.5°C/h) to 950°C in order to induce an efficient recrystallization of the material around the junction. Subsequently the system was cooled down at 200°/h up to room temperature. Oxygen annealing at 450°C during 200 h was applied afterwards.

The microstructural quality of the joined samples was analysed with a polarised-light optical microscope (Olympus Vanox AHMT3) and a SEM-EDX (Philips XL30 FEG-ESEM).

The superconducting properties were characterised by micro Hall Probe Mapping in liquid nitrogen. The Hall probe active area was 0.1 mm by 0.1 mm and the measurements were performed with a 0.5 mm step on a 25 mm x 25 mm scanned surface at 1 mm from the sample top surface. The sample is magnetised following a Field Cooling (FC) procedure using an air coil providing an applied magnetic field of $\mu_0H$ = 380 mT in liquid nitrogen. The $\rho(T)$ transport measurements were obtained using a conventional 4-point technique in a Quantum Design Physical Property Measurement System (PPMS) with applied magnetic fields, $\mu_0H$, ranging between 0 and 3 T. The resistivity measurements were performed following a Zero Field Cooling (ZFC) procedure in a range between 60 K and 100 K with a sweep rate of 0.3 K/min. The temperature step was 0.1 K and the applied current was 10 mA.

RESULTS AND DISCUSSIONS

A. Microstructural characterisations

SEM micrographs of welded samples are presented in figure 2. A junction along a plane perpendicular to the *c*-axis (usual configuration of junction) is shown in figure 2(A), and one in an *ab*-plane in figure 2(B). In both cases, it can be observed that the junctions are very clean. No secondary phase has been observed. Polarised light microscopy confirms that the junctions are well textured. The junction porosity is a little bit higher in the central part than near the edge of the single-domain. This porosity is probably due to impurities contained in the silver paint, to residual solvent traces, or to air trapping during the pressing.

In addition, by performing electronic microscopy in Back-Scattered Electron mode (BSE), no silver particle was found. Furthermore EDX measurements did not highlight the presence of silver in the ceramics near the junction. This might be due to the fact that sensitivity of EDX detection is however limited for light elements, such as silver, in a heavy matrix, here DyBaCuO. Nevertheless both results suggest that the silver content in the sample is very small, probably smaller than 1 to 2at%, and well distributed.

B. Electrical and magnetic properties

Both kinds of junctions were characterised by $\rho(T)$ measurements. Results for samples with the junction along a plane perpendicular to the *c*-axis are presented in figure 3, *i.e.*, the *c*-axis resistivity versus temperature curves with magnetic fields $\mu_0 H$ = 0, 1 and 3 T applied perpendicular to the *c*-axis. Intragrain, *G*, (filled symbols) and across the junction, *J*, (open symbols) measurements are presented on the same graph using a logarithmic scale for the resistivity axis.

For intragrain and across the junction measurements, the onset critical temperature, $T_c$, is about 92 K, which is a reasonable value for Dy-123. The resistive transitions within a grain are rather narrow whatever the applied magnetic field. For across the junction measurements, the transition is widened. Remarkably however, in the absence of magnetic field, the transition width is less than 4 K. Only a small shouldering is observed on $\rho(T)$ when measured across the junction. This behaviour is similar to that found by Doyle *et al.* [33,34]. This can be related to the junction weak link effect. Another reason for this shouldering in the $\rho$(T) curves might be related to macrocracks in the sample. Recall that in the studied configuration, the injected current flows along the *c*-axis (as shown on figure 2). This corresponds to a geometry in which eventual macrocracks, usually directed parallel to the *ab*-planes, may impede the superconducting current flow. Indeed such macrocracks have been observed by optical microscopy after $\rho$(T) measurements run between the $V_J$ contacts and not between the $V_G$ contacts. They may have induced a weak link effect responsible for a shouldering in the $\rho$(T) curve.

Samples with junction in a plane parallel to the *c*-axis were characterised by micro Hall probe mapping at T = 77 K (figure 4). The shape of the trapped field distribution exhibits one single peak and no decrease of the trapped field near the junction. Although Hall Probe mapping results should be interpreted with great care when analysing a single grain boundary [34], it is not possible to give evidence for a weak link effect of the junction by this technique.

Resistivity *vs* temperature measurements of samples with junction in a plane parallel to the *c*-axis are presented in figure 4. Intragranular (*G*) and across the junction (*J*) $\rho$(T) curve are plotted on the same graph, using a logarithmic scale for the resistivity axis. In this case, the current flows along the *ab*-planes in presence of magnetic fields $\mu_0 H°$ ranging from 0 to 3 T, and applied parallel to the *c*-axis and perpendicular to the injected current.

The transition is narrow and the onset critical temperature, $T_c$, is about 91 K. This $T_c$ value is slightly lower than the $T_c$ of samples with a junction perpendicular to the *c*-axis (cf. figure 2). This small difference can be attributed to a better oxygenation in the case of the sample with junction perpendicular to the *c*-axis.

The $\rho_G$(T) and $\rho_J$(T) curves are very similar and very close to each other for all applied magnetic fields. No intermediate shoulder is observed on $\rho_J$(T) curves, and these curves are only shifted by almost 0.2 K. Therefore the junction does not act as a weak link nor significantly affecting the critical current density.

It can be observed that the resistivity at T > $T_c$ is higher along the *c*-axis (figure 3) than in the *ab*-planes (figure 4). This can be related to the anisotropy of the Y-123-type structure.

CONCLUSIONS

Silver paint has been successfully used to weld Dy-Ba-Cu-O single-domains. Junctions have been performed in planes perpendicular or parallel to the *c*-axis. The microstructure of the junctions in both studied configurations is very clean, no secondary phase nor Ag particles can be observed. Optical microscopy allows concluding that the texturation occurs in all the material around the junction.

The junction parallel to the c-axis is characterized by better electrical properties than the junction perpendicular to the c-axis. For samples with the junction parallel to the c-axis, the narrow superconducting resistive transition, which is very close to the intragranular transition, shows that the junction does not alter significantly the superconducting properties of the samples.

ACKNOWLEDGEMENTS

The preparation and the characterization of materials are parts of the doctorate thesis of J-P Mathieu who thanks FRIA (Fonds pour la Formation à la Recherche dans l Industrie et dans l Agriculture, Brussels) for financial support. Part of this work results from research activities in the framework of the VESUVE project of the Walloon Region (RW.01.14881), and of the SUPERMACHINES project of the EC.

FIGURE CAPTIONS

**Figure 1:** Description of the junction configurations

**Figure 2:** SEM micrographs of the welded samples. (A) Junction along a (100) or (010) plane; (B) junction along a (001) plane. The observed plane is indicated by a dash line on the schematic representation of the sample.

**Figure 3:** Temperature dependence of the resistivity along the *c*-axis (junction perpendicular to the *c*-axis) with applied magnetic fields $\mu_0 H$ = 0, 1 and 3T. The field is perpendicular to the current flow and to the c-axis. Filled symbols correspond to intragrain measurements, whereas open symbols are for  across the junction  measurements. Note the extended log scale for the resistivity axis.

**Figure 4:** Hall Probe Mapping at 77 K of a welded sample with junction parallel to the c-axis. The arrow corresponds to the axe of the junction. A view of the sample has been plotted in order to correlate the trapped field with the surface probed by positioning precisely the X and Y axes.

**Figure 5:** Temperature dependence of the resistivity along *ab*-planes (junction parallel to the *c*-axis) with applied magnetic fields $\mu_0 H$ = 0, 1 and 3˚T. The field is perpendicular to the current flow and parallel to the *c*-axis. Filled symbols correspond to intragrain measurements, whereas open symbols are related to  across the junction  measurements. Note the log scale for the resistivity axis.

# *Figure 1*

Junction along (100) or (010) planes

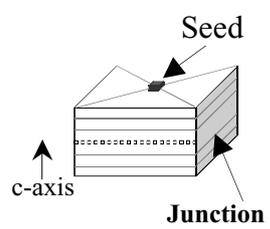

Junction along (001) planes

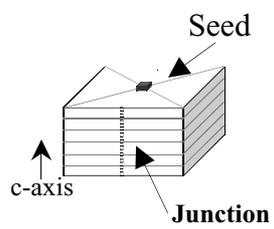

# *Figure 2*

Junction along a (100) or (010) plane

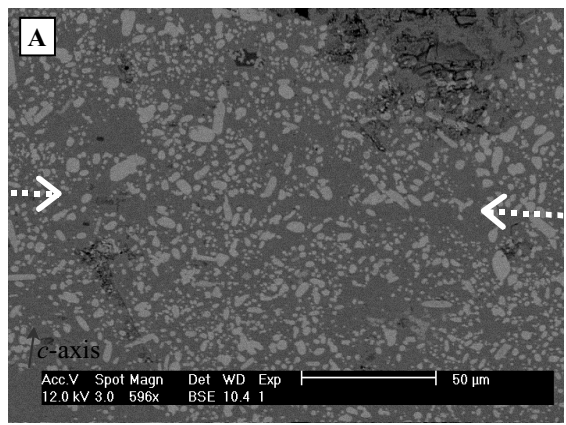
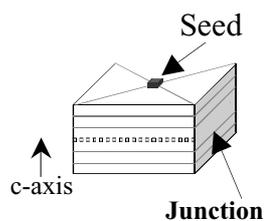

Junction along a (001) plane

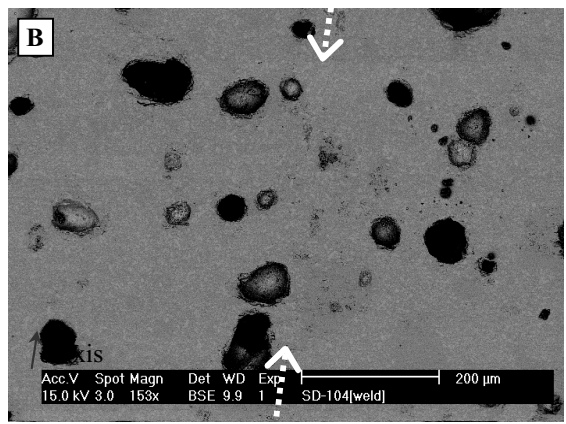
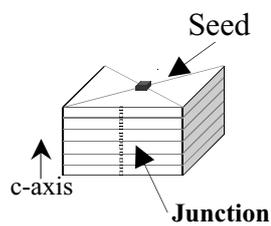

*Figure 3*

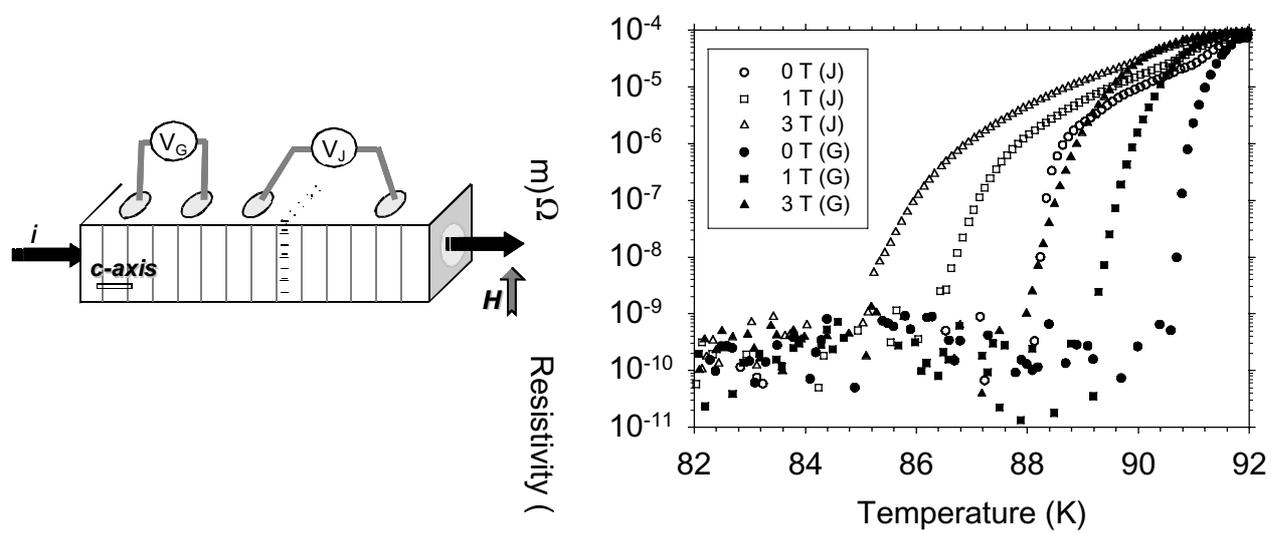

*Figure 4*

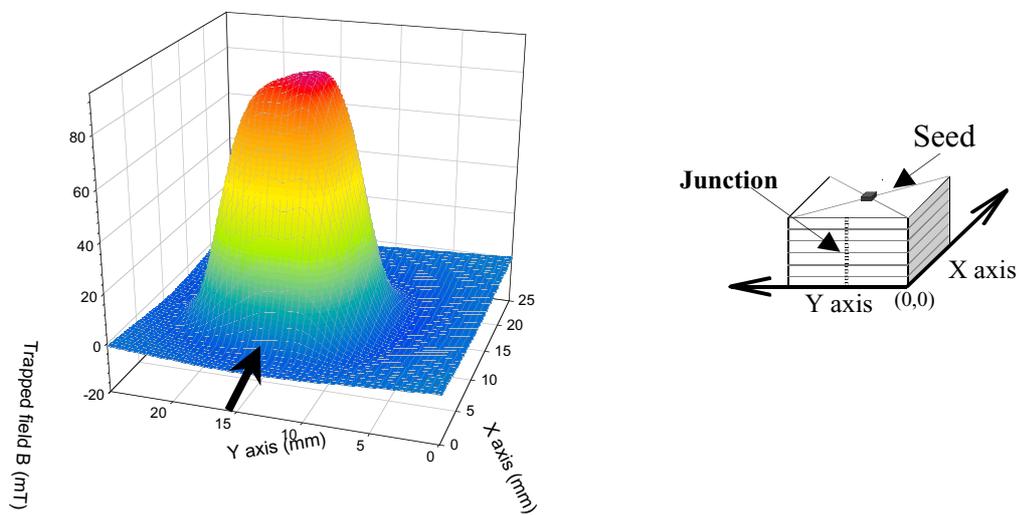

*Figure 5*

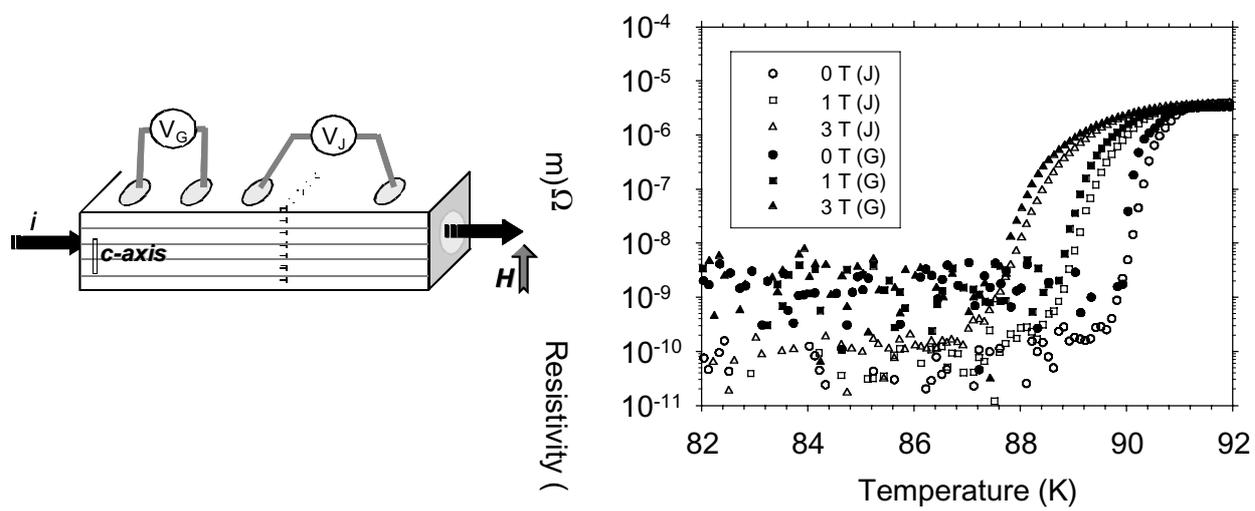